\documentstyle[prc,preprint,aps]{revtex}
\begin{document}
\title{First-forbidden $\beta$ decay of $^{17}$N and $^{17}$Ne}
\author { D.\ J.\ Millener\thanks{Permanent address: Brookhaven 
National Laboratory, Upton, New York 11973. Electronic address: 
millener@bnl.gov}}
\address{Brookhaven National Laboratory, Upton, New York 11973}
\address{and Institute for Nuclear Theory, University of Washington,
Seattle, Washington 98195}
\date{\today}
\maketitle
\begin{abstract}
 It is shown that differences, due to charge-dependent effects, in 
the $^{17}$N and $^{17}$Ne ground-state wave functions account for 
the fact that the experimentally measured branch for the $\beta^+$ 
decay of $^{17}$Ne to the first excited state of $^{17}$F is roughly
a factor of two larger than expected on the basis of nuclear matrix
elements which reproduce the corresponding $\beta^-$ branch in the
decay of $^{17}$N. 
\end{abstract}
\pacs{PACS numbers: 23.40.Hc, 21.60.Cs, 27.20.+n}

 By measuring positrons in coincidence with 495-keV $\gamma$ 
rays de-exciting the $1/2^+$ first-excited state of $^{17}$F,
Borge {\it et al.} \cite{bor93} have obtained a branch of 
1.65(16)\% for the first-forbidden $\beta^+$ decay of $^{17}$Ne
to the $1/2^+$ state. This is a very interesting result
because the measured branch is roughly a factor of two
larger than expected on  the basis of nuclear matrix elements
which reproduce the corresponding $\beta^-$ branch
of 3.0(5)\% \cite{pol73,alb76} in the decay of $^{17}$N.
Recently, Ozawa {\it et al.} \cite{oza96} have confirmed the
magnitude of the $\beta$ branch in $^{17}$Ne decay, obtaining
a value of 1.44(16)\% by a method which utilizes a 32 MeV/A
radioactive beam of $^{17}$Ne.

 The $\beta$-decay rate is given by $ft = 6170$ sec. For the 
$1/2^-\to 1/2^+$ transitions of interest,
$f = f^{(0)} + f^{(1)}$ where the superscript refers to the
spherical tensor rank of the  $\beta$-decay operator. In general,
$f^{(0)}$ is much larger than $f^{(1)}$ and, to a very good 
approximation,
\begin{equation}
f^{(0)} = I_0 (\xi'v +\frac{1}{3}W_0 w +\xi w')^2
\label{eq:f0}
\end{equation}
where $\xi = \pm \alpha Z/2R$ for $\beta^{\mp}$ decay, with $Z$ 
the charge of the daughter nucleus and $R=3.499$ fm for $A=17$, 
and
\begin{eqnarray}
w & = & \lambda\sqrt{3} {\hat J_f}/{\hat J_i}\langle J_f T_f||
ir[C_1 ,\sigma]{1\over\sqrt{2}}\tau ||J_i T_i\rangle C 
\label{eq:w} \\
\xi' v & = & -\lambda\sqrt{3} {\hat J_f}/{\hat J_i}\langle 
J_f T_f||{i\over M}
[\sigma ,\nabla ]{1\over\sqrt{2}}\tau ||J_i T_i\rangle C 
{\lambdabar_{Ce}^2} 
\label{eq:v}
\end{eqnarray}
with $C$ being the isospin Clebsch-Gordan coefficient and 
$\lambda = 1.26$. Energies are expressed in units of the 
electron rest mass and, with $I_0$ the integrated phase-space 
factor for allowed decays divided by the square of the Compton 
wave length for the electron, the nuclear matrix elements are 
in fm. The matrix element $w'$ is closely related to $w$ and 
takes a value $\sim 0.7w$ \cite{mil82}. These expessions are 
based on a  systematic expansion of the electron radial wave 
functions developed by Behrens and B\"{u}hring \cite{beh82}, 
the arcane notation for the nuclear matrix elements in
first-forbidden decays being historical (see \cite{mil82} for
details and definitions of the rank-1 matrix elements).

 Aside from the use of first-forbidden $\beta$ decay as a 
spectrocopic tool, there has been great interest in rank-0 
decays for two reasons. The first dates back to the suggestion 
\cite{kub78} that the matrix element $\xi' v$ of the time-like 
piece of the axial current $\gamma_5$ should be strongly 
enhanced by meson-exchange currents, largely one-pion-exchange.
This enhancement is now well established at $\sim 60$\% for 
light nuclei \cite{war94} and even larger for heavy nuclei 
\cite{war94b}. It is often taken into account, as is done below,
by multiplying $\xi' v$ by a factor $\varepsilon_{mec}$. The 
second reason relates to the similarity of the operators for 
parity-mixing and rank-0 first-forbidden $\beta$-decay 
\cite{ade85}. As a result of these fundamental interests, a 
large literature exists on many aspects of first-forbidden 
$\beta$ decay and parity-mixing in light nuclei. The present
treatment of the $^{17}$N and $^{17}$Ne decays, first studied
theoretically by Towner and Hardy \cite{tow72}, is based on 
a systematic study \cite{mil85} of $J^\pm \to J^\mp$ decays 
of $^{11}$Be, $^{15}$C, $^{16}$C, $^{16}$N, $^{17}$N, $^{18}$Ne,
$^{19}$Ne and $^{20}$F.

 For the $1\hbar\omega$ basis used in \cite{mil85}, the $1/2^-$
initial-state wave functions have a particularly simple form in
a weak-coupling representation, namely that of a $0p$-shell hole
coupled to $(1s0d)^2$ eigenstates (notation $J^\pi_n;T$).
\begin{eqnarray}
 |1/2^-;3/2\rangle &=& 0.967|1/2^-\otimes 0^+_1 ;1\rangle  
-0.224|3/2^-\otimes 2^+_1 ;1\rangle \nonumber \\
 & & +0.109|1/2^-\otimes 0^+_2 ;1\rangle + ...
\label{eq:wfn}
\end{eqnarray}
In fact, the three components listed account for 99.7\% of the 
wave function. For the  dominant component, only the $1s_{1/2}^2$
component contributes to the matrix element ${\bf \sigma .r}$ and
${\bf\sigma .p}$, one $s_{1/2}$ nucleon making a transition to 
fill the $p_{1/2}$ hole with the other forming the 
single-particle final state. The same is true for the third 
component, which augments the first (the $0^+_2;1$ state has a 
dominant $1s_{1/2}^2$ component). A small $d_{3/2}\to p_{3/2}$ 
amplitude, arising from the second component of the $1/2^-$ 
wave function, is important because the single-particle matrix 
element is large (larger than $s\to p$ by a factor of 
$\sqrt{5}$ for harmonic oscillator wave functions) and 
interferes destructively with the dominant $1s_{1/2}\to 
0p_{1/2}$ amplitude. This is a common feature of all the 
transitions studied in \cite{mil85}. The radial single-particle
matrix elements are computed with Woods-Saxon wave functions
obtained by adjusting the well depth to match the separation
energy from the initial or final state to the appropriate 
physical core states of the A-1 system \cite{mil85}. For the 
$1s_{1/2}\to 0p_{1/2}$ contribution, the only important parent 
states are the lowest $0^-$ and $1^-$ states of $^{16}$N or 
$^{16}$F. The separation energies are given in Table 
\ref{tab:energies} along with the decay energies and the 
phase-space integrals $I_0$. Since the 
separation energies are close to the Hartree-Fock
energies, the Woods-Saxon wave functions should be a good
approximation to one-nucleon overlap functions \cite{sat83}.

 For the rank-0 contribution to the $\beta$-decay rates, the
calculation gives
\begin{eqnarray}
f^{(0)}({\rm N}) & = & 0.3051(10.971\varepsilon_{mec}-4.216)^2 
\label{eq:fn} \\
f^{(0)}({\rm Ne}) & = & 2.380(11.585\varepsilon_{mec}-3.009)^2 
\label{eq:fne} \\
f^{(0)}({\rm Ne}') & = & 2.380(15.278\varepsilon_{mec}-3.969)^2
\label{eq:fnep} 
\end{eqnarray}
where the first two lines correspond to using identical nuclear
structure, the small differences in matrix elements being due
to the use of Woods-Saxon wave functions bound at the physical
separation energies (note the energy-dependent factors in Eq.
(\ref{eq:f0}) for the second term).  The resulting $f$ values 
are compared with experiment in Table \ref{tab:f} for two values
of the enhancement due to meson-exchange currents (see Table IV
of \cite{war94} for theoretical estimates of $\varepsilon_{mec}$). 
Including the calculated $f^{(1)}$ values, it can be seen that
the predicted value for the $\beta$ branch in $^{17}$Ne is less
than $\sim 0.9$\% for values of $\varepsilon_{mec}$ which produce
agreement with the $^{17}$N data (0.77\% to reproduce the central 
value).
 
 For the case denoted by Ne$'$ in Eq. (\ref{eq:fnep}) and the
last line of Table \ref{tab:f}, the $^{17}$Ne ground-state 
wave function has been modified to take into account 
charge-dependent effects which differ for $1s$ and $0d$ orbits. 
Now, with a $45-50$\% enhancement from meson-exchange currents, 
the calculated beta-decay rates are in agreement, within the 
error bars, for both nuclei.

 That there should be substantial $T_z$-dependent effects is 
evident from the 376 keV difference in Coulomb energies for 
the $0d_{5/2}$ and $1s_{1/2}$ orbits at $A=17$. For $A=18$, the 
large shift in the excitation energy of the the third $0^+$ 
state in $^{18}$Ne (Table \ref{tab:a18}) led to its 
identification as a largely $1s_{1/2}^2$ configuration 
\cite{ner74}. The shift in the $s_{1/2}^2$ diagonal matrix 
element relative to $d_{5/2}^2$ in going from $^{18}$O to 
$^{18}$Ne will also lead to more $s_{1/2}^2$ in the $^{18}$Ne 
ground-state wave function and hence, when coupled to a 
$p_{1/2}$ hole, to an enhancement of the 
rank-0 matrix element for the $\beta^+$ decay of $^{17}$Ne. 
This effect is amplified by the cancellation between the 
$s_{1/2}\to p_{1/2}$ and $d_{3/2}\to p_{3/2}$ contributions.

 To make a rough estimate of this effect, the Wildenthal USD 
interaction \cite{wil84} is used to obtain $(sd)^2$ wave 
functions for $^{18}$O ($\epsilon_{5/2}=-3.9478$, 
$\epsilon_{1/2}=-3.1635$,
$\epsilon_{3/2}=1.6466$, upper half diagonal of two-body
matrix elements -2.8197, -1.3247, -3.1856, -2.1246, -1.0835, 
-2.1845). Then, the $s_{1/2}^2$ diagonal matrix element is 
shifted by twice the shift of the $s_{1/2}$ single-particle 
energies between $^{17}$O and $^{17}$F (752 keV) plus 147 keV 
for the difference between the two-body matrix elements of 
$e^2/r$ for $d^2$ and $s^2$ configurations \cite{kah72}, and
the new matrix is diagonalized to get $(sd)^2$ wave functions 
for $^{18}$Ne. The resulting energies, wave functions, and 
intensities of $1s_{1/2}^2$ are given in Table \ref{tab:sd2}.
The $s_{1/2}^2$ intensity rises from 15\% to 21.7\%, an 
increase of 44\% (the squared overlap of the ground-state wave 
functions is still 0.9925). The increase in $\xi' v$ in Eq. 
(\ref{eq:fnep}) by a factor 1.32 rather than 1.20 for the 
$s_{1/2}\to p_{1/2}$ matrix element alone is due to the 
cancellation effects involving the $d_{3/2}\to p_{3/2}$ 
matrix element.

 The above calculation, which does succeed in providing an
explanation for the measured $\beta$-decay rates, is not a
consistent one but clearly indicates the direction in which
charge-dependent effects will affect the $\beta$-decay branch
in $^{17}$Ne decay. An explanation of the energy shifts and 
wave function changes for the $0^+$ $T=1$ states of $A=18$ 
requires that the $4p2h$ configurations be included. A 
calculation of the energy shifts without wave function changes 
\cite{ner74} does rather well but the $^{18}$Ne ground state 
could do with a ``push'' of the magnitude (163 keV) shown in 
Table \ref{tab:sd2}. The $(sd)^2$ calculation is actually 
more applicable to the $2p1h$ states of $^{17}$N and $^{17}$Ne 
because the $4p3h$ states are expected \cite{war89} to lie 
above both states obtained by coupling a $p_{1/2}$ hole to 
the two lowest $(sd)^2$ $0^+$ states. The second of these 
states is known at 3.663 MeV in $^{17}$Ne and is lowered 
from its position in $^{18}$O in large part because the 
spin-average $p_{1/2}^{-1}s_{1/2}$ $T=1$ particle-hole 
interaction is less repulsive by $\sim 700$ keV than the 
coresponding $p_{1/2}^{-1}d_{5/2}$ interaction
\cite{bar84,war89} and to a lesser extent because of the
removal of the influence of the $4p2h$ configuration. 

 To put the structure of $^{17}$N and $^{17}$Ne in a broader 
context, it should be noted that the four particle-hole matrix 
elements mentioned above can be deduced directly from the 
binding energies of the lowest four states of $^{16}$N (the
charge-dependent shifts of the $0d_{5/2}$ and $1s_{1/2}$
orbits, including a dependence on separation energy, can be seen
across these $T=1$ multiplets). Within the framework of the same
weak-coupling assumption used to deduce the particle-hole 
matrix elements, the total binding energies and multiplet 
spacings of the low-lying states of the heavy carbon and
nitrogen isotopes which contain one or more $sd$-shell
neutrons can be rather nicely accounted for (of course,
small components in the wave functions are important for
detailed spectroscopic applications such as first-forbidden
$\beta$ decay). In consistent 
shell-model calculations which include charge-dependent
interactions, the response to changes in $T_z$ on the one hand 
and to changes in the number of particles or holes on the other 
strongly restricts the $d_{5/2}/s_{1/2}$ content of the low-lying
states. An interesting case in the context of the present
study is $^{16}$C which has a rank-0 $\beta$ decay branch 
of 0.68\% \cite{gag83} to the lowest $0^-$ state of $^{16}$N.
With an extra $p_{1/2}$ proton hole, the energy of the excited 
$0^+$ state has been lowered to 3.02 MeV, implying slightly 
more $1s_{1/2}^2$ in the ground state than for $^{17}$N. The
first-forbidden $\beta$-decay rate is well accounted for using
the same type of shell-model calculation and meson-exchange
enhancement as for $^{17}$N \cite{mil85}.  

 A unique first-forbidden $\beta$ branch of 1.6(5)\% \cite{sil64}
to the ground state of $^{17}$O is known for the decay of $^{17}$N.
This branch corresponds to $f^{(2)} = 24(8)$. With no change in
the single nuclear matrix element involved, the expected branch
in $^{17}$Ne decay is 0.55(18)\%. Charge-dependent effects should
lower this value slightly because of a decrease in the $d_{5/2}^2$
component of the $^{17}$Ne ground state (Table \ref{tab:sd2}),
amplified somewhat by cancellation between $d_{5/2}\to p_{1/2}$
and $d_{5/2}\to p_{3/2}$ contributions. Shell-model calculations
with the basis of Ref. \cite{mil85} overpredict $f^{(2)}$ by
a little more than a factor of two for either harmonic oscillator
or Woods-Saxon wave functions. This is quite consistent with
a similar overestimate for the unique first-forbidden decay of
$^{16}$N for a correspondingly small shell-model basis. This
problem is resolved in calculations using a very large 
shell-model basis with all configurations up to $4\hbar\omega$
\cite{war94,war92}. The rank-0 matrix elements are also reduced in 
such calculations \cite{war94} but by a lesser amount due to a
cancellation between contributions from $2p2h$ admixtures induced
by central and tensor forces. The experimental $\beta$-decay
rates can then be reproduced using values for $\varepsilon_{mec}$
close to the theoretical value of about 1.6 \cite{war94}.

 In conclusion, the use of realistic (e.g., Woods-Saxon) radial 
wave functions is essential for evaluating first-forbidden 
$\beta$-decay matrix elements \cite{war94,mil85}, particularly 
for $1s_{1/2}\leftrightarrow 0p_{1/2}$ transitions for which 
the $1s_{1/2}$ nucleon is loosely bound, as is the case for 
the decay of $^{17}$N and $^{17}$Ne to the first-excited 
$1/2^+$ states of $^{17}$O and $^{17}$F. However, radial wave 
function differences do not account for the strong asymmetry 
observed for these decays. Rather, plausible $T_z$-dependent 
differences in the $1s_{1/2}$ occupancy for the initial states
can account for the asymmetry. Furthermore, the very 
small separation energy for the $1s_{1/2}$ proton in $^{17}$F is 
not germane to the problem since this proton is a spectator in 
the $\beta$-decay process. In fact, from the way in which the
parentage expansion is made and separation energies determined,
the spectator $1s_{1/2}$ proton forms part of a $^{16}$F core
where it is unbound for the physical core states (by 535 keV for
the $0^-$ state). Substantial asymmetries have also been observed
for the allowed decays of $^{17}$N and $^{17}$Ne \cite{til93}.
While overlap factors for radial wave functions bound at different
energies now play a role because the Gamow-Teller operator has no 
spatial structure, it again seems likely that the observed 
asymmetries are largely due to $T_z$-dependent mixing of various 
shell-model configurations. For the $2p1h$ configurations with 
$T=1/2$, the mixing of configurations with $T=0$ and $T=1$ for the
$(sd)^2$ configurations determines both the overall spatial 
symmetry and the relative contributions to the Coulomb energy from
$p$ and $sd$ orbits. There are also low-lying $4p3h$ configurations 
(one $1/2^-$ and two $3/2^-$) which have their own Coulomb 
energy shifts and mix strongly with the $2p1h$ configurations.
Thus, there should be significant $T_z$-dependent mixing in 
both the initial and final states for the Gamow-Teller decays.
A beautiful demonstration of this type of $T_z$-dependent mixing
is seen in changes of the ratio of Gamow-Teller strengths for the
lowest two $2^+;T=1$ states reached via $(n,p)$, $(p,p')$ and 
$(p,n)$ reactions on $^{14}$N \cite{jackson}. Here, the near 
degeneracy of $2h$ and $2p4h$ configurations \cite{war60}, with
Coulomb energies that differ by $\sim 700$ keV across the 
multiplet, leads to very different wave functions for each nucleus.

\vspace{\baselineskip}
This research was supported by the U. S. Department of Energy 
under Contract No. DE-AC02-76CH00016 with Brookhaven National 
Laboratory.


\narrowtext
\squeezetable
\begin{table}
\caption[a]{Parameters governing the decays of $^{17}$N and 
$^{17}$Ne to the first-excited states of $^{17}$O  and $^{17}$F. 
Separation energies are given for the $0^-_1;1$ core states in 
$^{16}$N and $^{16}$F; the values for the $1^-_1$ core state are
0.28 MeV and 0.19 MeV higher, respectively.}
\begin{tabular}{ccccc}
 & $W_0$ & $I_0$ & $S_{n/p}(s_{1/2})$ & $S_{p/n}(p_{1/2})$ \\
 & (MeV) &       & (MeV)              & (MeV) \\
\tableline
 $^{17}$N & ~8.32 MeV & 0.3051 & 6.00 ($n$) & 13.03 ($p$) \\
 $^{17}$Ne & 13.52 MeV & 2.380 & 1.48 ($p$) & 16.80 ($n$) \\
\end{tabular}
\label{tab:energies}
\end{table}

\narrowtext
\squeezetable
\begin{table}
\caption[b]{Comparison of theoretical and experimental 
$\beta$-decay  rates via $f$ values. $f_{exp}$ for $^{17}$Ne
decay is derived from the average 1.55(12)\% of the two
measurements \cite{bor93,oza96} for the $\beta$ branch.}
\begin{tabular}{ccccc}
 & \multicolumn{2}{c}{$f^{(0)}$} & $f^{(1)}$ & $f_{exp}$  \\
 $\varepsilon_{mec}$ & 1.4 & 1.5  & & \\
\tableline
 $^{17}$N & 37.9 & 45.7 & 6.5 & 44.4(74) \\
 $^{17}$Ne & 415 & 491 & 21 & 873(64) \\
 $^{17}$Ne$'$ & 722 & 854 & 21 & 873(64) \\
\end{tabular}
\label{tab:f}
\end{table}

\narrowtext
\squeezetable
\begin{table}
\caption[c]{Excitation energies (MeV) of $0^+$ $T=1$ states
relative to the lowest such state. The $0^+_2$ states are mainly
$4p2h$ in nature. In the case of $^{18}$F, it should be noted 
that the lowest $0^+$ state obtains extra binding energy from 
the charge-independence breaking $np$ interaction \cite{kah72}.}
\begin{tabular}{cccc}
 $J_n^\pi$ & $^{18}$O & $^{18}$F & $^{18}$Ne \\
\tableline
 $0^+_3$ & 5.336 & 5.094 & 4.590 \\
 $0^+_2$ & 3.630 & 3.711 & 3.576 \\
\end{tabular}
\label{tab:a18}
\end{table}

\narrowtext
\squeezetable
\begin{table}
\caption[d]{Results of $(sd)^2$ diagonalizations. Wave function
amplitudes are given in columns $4-6$. The binding energy of
the $0^+_1$ state of $^{18}$O is chosen as the zero of energy.}
\begin{tabular}{ccccccc}
 & $J_n^\pi$ & $E_x$ & $d_{5/2}^2$ & $s_{1/2}^2$ & $d_{3/2}^2$ & 
 \%$s_{1/2}^2$ \\
\tableline
 $^{18}$O & $0^+_1$ & ~0.000 & 0.8886 & ~0.3878 & 0.2448 & 15.0 \\
          & $0^+_2$ & ~4.320  & 0.3932 & $-0.9190$ & 0.0287 & 84.5 \\
 $^{18}$Ne & $0^+_1$ & $-0.163$ & 0.8521 & ~0.4654 & 0.2394 & 21.7 \\
          & $0^+_2$ & ~3.588  & 0.4667 & $-0.8827$ & 0.0547 & 77.9 \\
\end{tabular}
\label{tab:sd2}
\end{table}

\end{document}